\documentclass[twocolumn,tighten]{aastex63}

\usepackage{xcolor}
\definecolor{my_color}{HTML}{3a18b1}

\definecolor{new_color}{HTML}{CF0000}
\definecolor{new_black}{HTML}{000000}
\usepackage{xfrac}

\newcommand\bedit[1]{\textcolor{new_black}{{#1}}}

\newcommand{\Kepler}{{\it Kepler}}

\newcommand{\Gaia}{{\it Gaia}}

\newcommand{\vespa}{\texttt{vespa}}

\newcommand{\be}{\begin{equation}}
\newcommand{\ee}{\end{equation}}

\newcommand{\metallicity}{[M/H]}
\newcommand{\msun}{M\ensuremath{_\odot}}
\newcommand{\rsun}{R\ensuremath{_\odot}}

\newcommand{\gcc}{\ensuremath{\rm g\,cm^{-3}}}

\newcommand{\mearth}{M$_\oplus$}
\newcommand{\rearth}{R$_\oplus$}

\newcommand{\lumsb}{0.00533}
\newcommand{\ulumsb}{0.00046}
\newcommand{\lum}{0.00516}
\newcommand{\ulum}{0.00020}

\newcommand{\qone}{0.52}
\newcommand{\uqone}{0.14}
\newcommand{\qtwo}{0.229}
\newcommand{\uqtwo}{0.036}
\newcommand{\uone}{0.330}
\newcommand{\uuone}{0.070}
\newcommand{\utwo}{0.392}
\newcommand{\uutwo}{0.069}
\newcommand{\offset}{0.000030}
\newcommand{\uoffset}{0.000029}
\newcommand{\rhost}{22.5}
\newcommand{\urhost}{1.4}
\newcommand{\rprstb}{0.0402}
\newcommand{\urprstb}{0.0018}
\newcommand{\arstb}{44.77}
\newcommand{\uarstb}{0.93}
\newcommand{\inclb}{89.15}
\newcommand{\uinclb}{\ensuremath{^{+0.11}_{-0.079}}}
\newcommand{\impb}{0.65}
\newcommand{\uimpb}{\ensuremath{^{+0.072}_{-0.12}}}
\newcommand{\rplb}{1.017}
\newcommand{\urplb}{0.051}
\newcommand{\perplb}{8.689099}
\newcommand{\uperplb}{0.000025}
\newcommand{\ttransitb}{2455374.6219}
\newcommand{\uttransitb}{0.0016}
\newcommand{\rprstc}{0.042}
\newcommand{\urprstc}{\ensuremath{^{+0.0055}_{-0.0038}}}
\newcommand{\arstc}{76.8}
\newcommand{\uarstc}{1.6}
\newcommand{\inclc}{89.339}
\newcommand{\uinclc}{0.056}
\newcommand{\impc}{0.875}
\newcommand{\uimpc}{0.074}
\newcommand{\rplc}{1.06}
\newcommand{\urplc}{\ensuremath{^{+0.15}_{-0.10}}}
\newcommand{\perplc}{19.53527}
\newcommand{\uperplc}{0.00010}
\newcommand{\ttransitc}{2455410.9777}
\newcommand{\uttransitc}{0.0033}

\newcommand{\mst}{0.1977}
\newcommand{\umst}{0.0051}
\newcommand{\rst}{0.2317}
\newcommand{\urst}{0.0049}
\newcommand{\tdurb}{1.184}
\newcommand{\utdurb}{\ensuremath{^{+0.085}_{-0.066}}}
\newcommand{\tdurc}{1.07}
\newcommand{\utdurc}{\ensuremath{^{+0.15}_{-0.21}}}
\newcommand{\teqb}{307}
\newcommand{\uteqb}{26}
\newcommand{\teqc}{234}
\newcommand{\uteqc}{20}
\newcommand{\fluxb}{2.208}
\newcommand{\ufluxb}{0.094}
\newcommand{\fluxc}{0.750}
\newcommand{\ufluxc}{0.032}

\newcommand{\thisstar}{{\em Kepler}-1649}
\newcommand{\thisfirstplanet}{{\em Kepler}-1649~b}
\newcommand{\thissecondplanet}{{\em Kepler}-1649~c}

\submitjournal{ApJL}

\shorttitle{The \Kepler\ Planetary Rescue Committee}
\shortauthors{Vanderburg et al.}

\begin{document}

\title{A Habitable-Zone Earth-Sized Planet Rescued from False Positive Status}

\author[0000-0001-7246-5438]{Andrew Vanderburg}
\altaffiliation{NASA Sagan Fellow}
\affiliation{Department of Astronomy, The University of Texas at Austin, Austin, TX 78712, USA}

\author[0000-0002-4829-7101]{Pamela~Rowden}
\affiliation{School of Physical Sciences, The Open University, Milton Keynes MK7 6AA, UK}

\author[0000-0003-0081-1797]{Steve Bryson}
\affiliation{NASA Ames Research Center, Moffett Field, CA 94901, USA}

\author[0000-0003-1634-9672]{Jeffrey Coughlin}
\affiliation{SETI Institute, 189 Bernardo Ave, Suite 200, Mountain View, CA 94043, USA}

\author[0000-0002-7030-9519]{Natalie Batalha}
\affiliation{University of California Santa Cruz, Santa Cruz, CA, USA}

\author[0000-0001-6588-9574]{Karen A.\ Collins}
\affiliation{Center for Astrophysics \textbar \ Harvard \& Smithsonian, 60 Garden Street, Cambridge, MA 02138, USA}

\author[0000-0001-9911-7388]{David W. Latham}
\affiliation{Center for Astrophysics \textbar \ Harvard \& Smithsonian, 60 Garden Street, Cambridge, MA 02138, USA}

\author[0000-0001-7106-4683]{Susan E. Mullally}
\affiliation{Space Telescope Science Institute, Baltimore, MD 21218, USA}

\author[0000-0001-8020-7121]{Knicole D. Col\'{o}n}
\affiliation{NASA Goddard Space Flight Center, Exoplanets and Stellar Astrophysics Laboratory (Code 667), Greenbelt, MD 20771, USA}	

\author{Chris Henze}
\affiliation{NASA Ames Research Center, Moffett Field, CA 94901, USA}

\author[0000-0003-0918-7484]{Chelsea X. Huang}
\newcommand{\MIT}{Department of Physics and Kavli Institute for Astrophysics and Space Research, Massachusetts Institute of Technology, Cambridge, MA 02139, USA}

\author[0000-0002-8964-8377]{Samuel N. Quinn}
\affiliation{Center for Astrophysics \textbar \ Harvard \& Smithsonian, Cambridge, MA 02138, USA}

\correspondingauthor{Andrew Vanderburg}
\email{avanderburg@utexas.edu}



\begin{abstract}

We report the discovery of an Earth-sized planet in the habitable zone of a low-mass star called \thisstar. The planet, \thissecondplanet, is \rplc\urplc\ times the size of Earth and transits its \mst\ $\pm$ \umst\ \msun\ mid M-dwarf host star every 19.5 days. It receives 74 $\pm$ 3 \% the incident flux of Earth, giving it an equilibrium temperature of \teqc\ $\pm$ \uteqc K and placing it firmly inside the circumstellar habitable zone. \thisstar\ also hosts a previously-known inner planet that orbits every 8.7 days and is roughly equivalent to Venus in size and incident flux. \thissecondplanet\ was originally classified as a false positive by the \Kepler\ pipeline, but was rescued as part of a systematic visual inspection of all automatically dispositioned \Kepler\ false positives. This discovery highlights the value of human inspection of planet candidates even as automated techniques improve, and hints that terrestrial planets around mid to late M-dwarfs may be more common than those around more massive stars. 

\end{abstract}



\section{Introduction} \label{sec:intro}

M-dwarf stars ($0.1 \msun \lesssim M \lesssim 0.6 \msun$)  are the most common outcome of the star formation process in our galaxy \citep{kroupaimf} --- they make up two-thirds of all stars and brown dwarfs in the solar neighborhood \citep{henry}.\footnote{\url{http://www.recons.org/census.posted.htm}} Exoplanet surveys have found that these stars frequently host terrestrial-sized planets, including those in temperate orbits which could possibly support liquid water on their surfaces \citep{dc13}.  Although M-dwarfs may be less hospitable than higher mass stars like the Sun \citep[e.g.,][]{lugerbarnes, wardhoward}, their sheer numbers make it plausible that planets around M-dwarfs may be the most common habitable environments in the universe. 

Our knowledge of planets around M-dwarfs has greatly increased in the last decade thanks to observations from the \Kepler\ space telescope. \Kepler\ was designed to measure how frequently planets are found around Sun-like stars, but a few thousand M-dwarf stars were also observed in its survey. From these data, we have learned that on average, each M-dwarf star hosts more than two sub-Neptune-sized planets with periods shorter than 200 days \citep{dc15}. Small planets are found more frequently around M-dwarfs than around higher-mass Sun-like stars \citep{mulders} and there is tentative evidence that this trend holds even within the M-dwarf spectral class: lower mass ($M_\star\approx 0.25 \msun$) ``mid'' M-dwarfs may host even more small planets than higher mass ($M_\star\approx 0.5 \msun$) ``early'' M-dwarfs  \citep{muirhead2015, hardegreeullman}. 

These and other statistical results have been enabled by the well-characterized planet candidate catalogs produced by the \Kepler\ mission.  Pixel time series were downloaded from the spacecraft and processed by the \Kepler\ pipeline, which performed image calibration, extracted light curves, removed instrumental systematics, and searched for periodic flux decrements that could be due to a transiting planet \citep{jenkins,Jenkins2017}; these initial detections are known as Threshold Crossing Events (TCEs).  The TCE Review Team (TCERT) then reviewed and dispositioned (classified) TCEs as either planet candidates (PCs) or false positives (FPs), with TCEs potentially due to any astrophysical transiting/eclipsing object given a Kepler Object of Interest (KOI) number.  By the end of the mission, the TCERT process was fully automated via the Robovetter \citep{coughlin2016,thompson}, which uses dozens of specialized metrics and a sophisticated decision tree to classify TCEs --- it only used Kepler observations, since other measurements (e.g., ground-based spectroscopic follow-up) were not uniformly performed on all targets.  This uniform vetting, along with associated synthetic data products \citep{Burke2017,Christiansen2017,Coughlin2017}, allows for the measurement of the final catalogs' completeness and reliability, thus enabling the accurate determination of planetary occurrence rates.  As a result, individual disposition correctness was sacrificed for statistical uniformity, and so it was known that some individual KOIs were incorrectly vetted, with interesting planets misclassified as false positives, and vice versa.



False positive KOIs are generally not followed up, possibly ignoring true planets that were incorrectly dispositioned as false positives.  To address this issue, members of our team formed the \Kepler\ False Positive Working Group (FPWG, \citealt{fpwg}) to visually inspect, using all available data, all \Kepler\ Objects of Interest (KOIs) classified as false positives by the Robovetter. Its goals were to:

\begin{enumerate}
    \item Produce a list of known ``certified'' false positives\footnote{\url{https://exoplanetarchive.ipac.caltech.edu/cgi-bin/TblView/nph-tblView?app=ExoTbls&config=fpwg}}, which can be used as a ground-truth when testing new classifiers. 
    \item Diagnose any issues or failure modes in the Robovetter's algorithm which could be corrected to improve its classifications. 
    \item Identify and rescue any viable planet candidates which were erroneously classified as false positives. 
\end{enumerate}

Over the past five years, the FPWG has inspected nearly 5,000 KOIs and certified nearly 4,000 KOIs as false positives or false alarms.  In the course of this review, we also examined objects that we were unable to certify as false positives and identified the ones most likely to be viable planet candidates.  Most recently, the FPWG finished a review of all Data Release 25 (DR25) FP KOIs, including those at very low signal-to-noise.  Among these possible planet candidates, one signal stood out as both particularly high quality and scientifically interesting: an Earth-sized planet candidate in a temperate orbit around a nearby low-mass star.

Here, we report our investigation of this newly rescued planet candidate signal. We take advantage of the fact that the candidate's host star, \thisstar, was already shown to be an exoplanet host by \citet{angelo}, who characterized and validated an inner planet called \thisfirstplanet. \citet{angelo} described \thisfirstplanet\ as a ``Venus analog'' because it is similar in size and incident flux to our Solar System neighbor. In this letter, we validate our newly rescued candidate as a planet in the system and show that in addition to a Venus analog, \thisstar\ hosts an Earth analog as well. Section \ref{observations} describes the observations and analysis we used to characterize this new signal and Section \ref{validation} describes our statistical validation of \thissecondplanet. Finally, we conclude in Section \ref{discussion} by discussing \thissecondplanet's characteristics, the system's architecture, and the implications of this detection regarding the occurrence rate of rocky, habitable planets around M-dwarfs.

\section{Observations and Analysis}\label{observations}

\subsection{\Kepler\ Light Curve}\label{lightcurve}


The \Kepler\ space telescope observed \thisstar\ (KIC~6444896 / KOI 3138) for a total of 756 days between 2010 and 2013 during its primary mission. \thisstar\ was observed during Quarters 6--9 (as part of guest investigator proposal GO20031, PI Di Stefano) and Quarters 12--17 (after KOI~3138.01/\thisfirstplanet\ was designated a planet candidate). 

A signal with period P~$\approx$~8.689~d was first detected, designated as KOI~3138.01, and dispositioned as a planet candidate in the \citet{burke14} catalog, which used data from Quarter 1 to Quarter 8, or only nine months of data for this target (Quarters 6, 7, and 8).  It was also dispositioned as a planet candidate in each subsequent catalog that re-examined it (the last two of which were based on the Robovetter \citealt{Coughlin2017,thompson}). This signal eventually became known as \thisfirstplanet\ after statistical validation by \citet{angelo}.

A second signal with P~$\approx$~19.535~d was only detected in the final \Kepler\ pipeline run \citep[TCE 6444896-02;][]{Twicken2016}.  The \citet{thompson} Robovetter dispositioned the TCE as a not-transit-like false positive (i.e., a false alarm) with a comment of ``\texttt{MOD\_NONUNIQ\_ALT}'', which indicates it failed the Model-Shift Uniqueness test \citep[see \S A.3.4 of][]{thompson}. This outcome indicates that the Robovetter judged the signal's significance to be too low compared to the systematic noise level to be a planet candidate.  The Robovetter assigned the 19.5~d TCE a ``disposition score'' \citep[see \S 3.2 of][]{thompson} of 0.374, which indicates only weak confidence in the false positive disposition (scores near 0.0 indicate high-confidence FPs, while scores near 1.0 indicate high-confidence PCs).  All TCEs with disposition scores $>$0.1 were assigned KOI numbers by \citet{thompson}, and thus KOI~3838.02 was created. For convenience, we refer to this new planet candidate (KOI~3138.02) as \thissecondplanet\ hereafter.


Our team inspected the transit signal of \thissecondplanet\ as part of our systematic review of \Kepler's false positive objects of interest. Though our assessments of the false positives were usually in good agreement with the Robovetter, we were not confident in the Robovetter's False Positive classification. Unlike most of the false alarms the Robovetter fails with the Model-Shift Uniqueness test, \thissecondplanet\ has a relatively short orbital period and dozens of observed transits, which were consistent in shape and depth over time. We found no compelling evidence that \thissecondplanet\ was a false positive, and instead identified it as a possible planet candidate.

We hypothesized that \thissecondplanet\ failed the Robovetter's Model-Shift Uniqueness test because of the atypical noise properties of the light curve produced by the \Kepler\ pipeline. The \Kepler\ pipeline light curve shows strong quarterly variations in its photometric scatter; for example, the photometric scatter in Quarter 7 ($\approx$3500 ppm) is almost four times greater than the photometric scatter in Quarter 6 ($\approx$900 ppm). We traced this effect to the choice of photometric apertures by the \Kepler\ pipeline. \thisstar\ has a high proper motion of 168.2 mas per year, but its motion was not taken into account when the \Kepler\ pipeline selected optimal pixels for aperture photometry \citep{pixelSelection}. Instead, the pixel selection algorithm assumed \thisstar's J2000 position, about 2 arcseconds (or half of a \Kepler\ pixel) from its true position during \Kepler's observations.  Because \thisstar\ is faint, its optimal photometric apertures were small enough that this half-pixel error caused \thisstar\ to fall at the edge or outside of the aperture in some quarters, while remaining within the photometric aperture in other quarters. The variation of the pixel position of \thisstar\ relative to the pixels selected for photometry significantly weakened the transit signal in the original pipeline photometry.

\begin{figure*}[htb] 
   \centering
   \includegraphics[width=\linewidth]{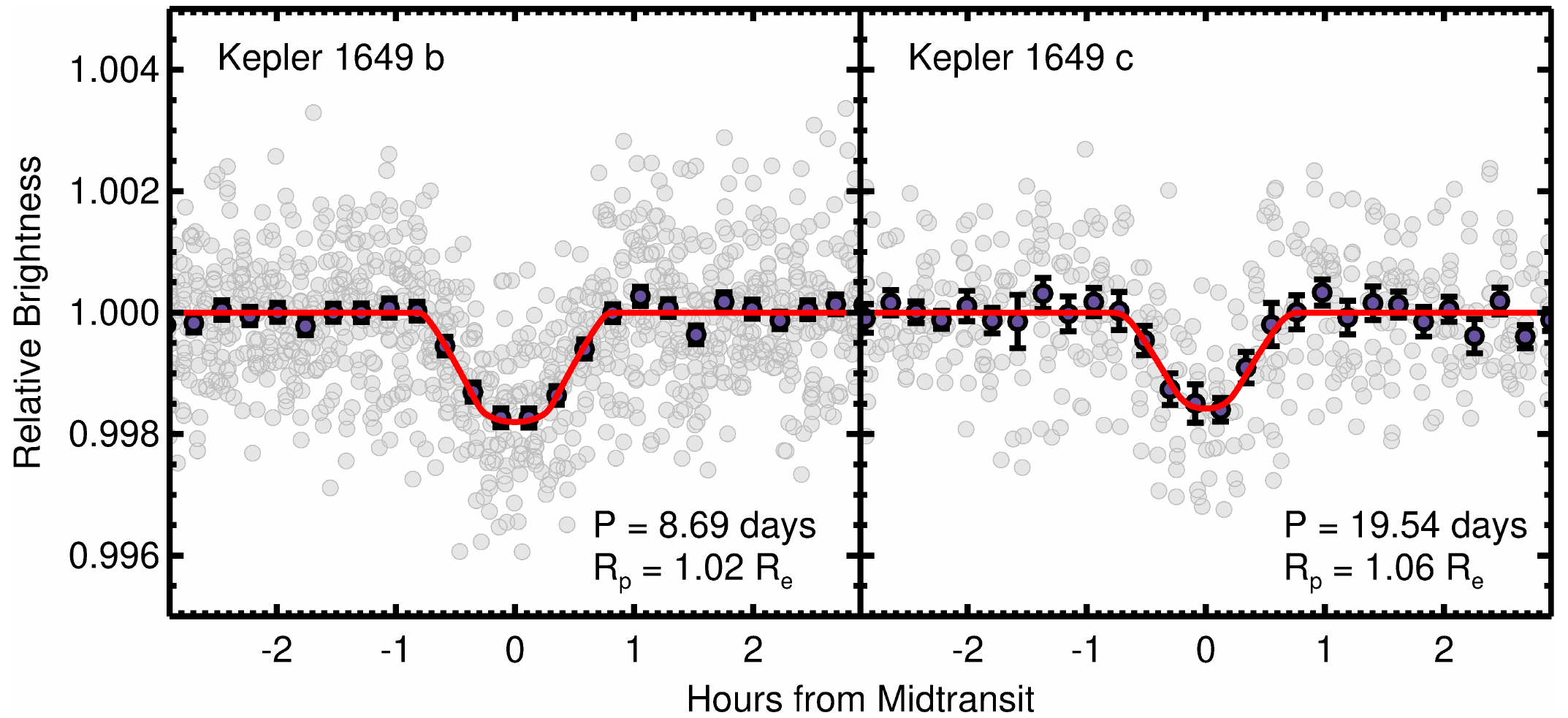} 
   \caption{\Kepler\ light curve phase-folded on the transits of \thisstar\ b (\textit{left}) and c (\textit{right}). Grey points are individual \Kepler\ long-cadence exposures, purple points are averages in phase, and the red solid line is the best-fit transit model.}
   \label{lc}
\end{figure*}

We therefore chose to produce our own light curves from the \Kepler\ target pixel files (Data Release 25). Following \citet{v15}, we extracted light curves from a set of 20 different photometric apertures. Half of these apertures were circular, defined by identifying all pixels within ten different radii of \thisstar's position. The other half were shaped like the \Kepler\ Pixel Response Function (PRF), defined by fitting the PRF to a representative \Kepler\ image for each quarter and selecting all pixels where the model PRF exceeded 10 different fractions of the peak model flux. We selected the aperture which produced the light curve with the highest photometric precision\bedit{\footnote{\bedit{We measured the photometric precision on 6 hour timescales by applying a 13 point standard deviation filter to the light curve and taking the median value. This is a quick approximation to the \Kepler\ pipeline's Combined Differential Photometric Precision (CDPP) metric \citep{cdpp}. For more information, see the documentation for the PyKE routine \texttt{kepstddev} (\url{https://keplerscience.arc.nasa.gov/ContributedSoftwareKepstddev.shtml}, \citealt{pyke}) and Section 3.1 of \citet{vj14}.}}} in each quarter after accounting for diluting flux from nearby fainter stars. The resulting light curve had much more consistent photometric scatter across quarters, improving significantly upon the \Kepler\ pipeline light curve in the quarters where its aperture selection was suboptimal. We use this new light curve in our analysis throughout the rest of the paper and show the phase-folded transit signals of \thisstar\ b and c in Figure \ref{lc}.  We also experimented with fitting the \Kepler\ images with a model based on the telescope's measured PRF to produce light curves (as done by \citealt{angelo}), but never achieved higher photometric precision than our well-optimized aperture photometry.

Finally, we re-ran the Model-Shift Uniqueness test using our new light curve, and found that this time, the planet candidate solidly passed. We therefore consider \thissecondplanet\ to be a viable planet candidate.

\subsection{Stellar Parameters}\label{stellarparameters}

Although the star \thisstar\ was already well characterized by \citet{angelo}, we were able to improve upon their stellar parameters thanks to the newly-released trigonometric parallax from \Gaia\ DR2 \citep{gaiamission, gaiadr2}. We derive \thisstar's radius and mass using empirical relations between these quantities and the star's absolute K-band magnitude from \citet{mann2015} and \citet{mann2019}, respectively. These relations yield a mass of $M_\star =$ \mst\ $\pm$ \umst\ \msun\ and radius of $R_\star =$ \rst\ $\pm$ \urst\ \rsun, which are consistent with, but several times more precise than, the estimates from \citet{angelo}. We adopt the spectroscopic metallicity and temperature reported by \citet{angelo} of [M/H] = -0.15 $\pm$ 0.11 and $T_{\rm eff} =$ 3240 $\pm$ 61 K. Using an average of three different bolometric corrections for the 2MASS $JHK$ magnitudes from \citet{mann2015}, we find that \thisstar\ is only about half a percent as luminous as the Sun ($L_\star$ = \lum\ $\pm$ \ulum\ $L_\odot$). This value is in good agreement with the luminosity calculated from our adopted effective temperature and stellar radius using the Stefan-Boltzmann law ($L_\star$ = \lumsb\ $\pm$ \ulumsb\ $L_\odot$), indicating that our stellar parameters are self-consistent. Our adopted stellar parameters are listed in Table \ref{bigtable}.

\subsection{Transit Modeling}\label{lcmodel}

We determined planetary parameters with a Markov Chain Monte Carlo (MCMC) analysis of the \Kepler\ light curve. We modeled the transit profiles of both planets simultaneously using \citet{mandelagol} curves, oversampling by a factor of 30 and integrating to account for the 29.4 minute \Kepler\ long-cadence exposure times. We fit for the host's stellar density, $\rho_\star$, imposing a Gaussian prior centered at \rhost\ \gcc\ with width \urhost\ \gcc, and calculated the scaled semimajor axes ($a/R_\star$) for both planets from the $\rho_\star$ value at each MCMC step using Kepler's third law. Because both \thisstar\ b and c orbit too far from their host star to have undergone tidal circularization, we model eccentric orbits for both planets, fitting in $\sqrt{e}\sin{\omega}$ and $\sqrt{e}\cos{\omega}$, where $e$ is the orbital eccentricity, and $\omega$ is the argument of periastron. Several groups have shown that planets in multi-transiting systems tend to have lower orbital eccentricity than planets in single-transiting systems, so we imposed a Gaussian prior on each planet's eccentricity centered at 0 with width 0.103, the 2$\sigma$ upper limit from \citet{vaneylen}. We assumed a quadratic limb darkening law, fitting for coefficients using the $q_1$ and $q_2$ parameterization recommended by \citet{kippingld}. Since the transits of \thisstar\ b and c are both short-duration and heavily distorted by the \Kepler\ long-cadence integration time, we imposed Gaussian priors on \thisstar's $u_1$ and $u_2$ limb darkening coefficients. The priors were centered on values ($u_1=0.33$, $u_2=0.39$) predicted by model atmospheres calculated by \citet{claretbloemen} and had widths of 0.07, the empirically measured scatter between these model predictions and measured values \citep{muller}. Finally, we limited the planet radii by enforcing $R_p/R_\star < 1$. In other words, the planets cannot be larger than their host star.

All in all, we fit 16 parameters: the stellar density, two limb darkening parameters ($q_1$ and $q_2$), a constant flux offset, and for each planet, orbital period, transit time, orbital inclination, $\log{(R_p/R_\star)}$, $\sqrt{e}\sin{\omega}$, and $\sqrt{e}\cos{\omega}$. We explored parameter space with an affine invariant MCMC sampler \citep{goodman} evolving each of 100 walkers for 100,000 steps, removing the first 10,000 steps as burn-in. The results of our fit are given in Table \ref{bigtable} and our best-fit model is plotted in Figure \ref{lc}.

\begin{deluxetable*}{lcc}[htbp]
\tabletypesize{\footnotesize}
\tablecaption{System Parameters for \thisstar \label{bigtable}}
\tablewidth{0pt}
\tablehead{
  \colhead{Parameter} & 
  \colhead{Value}     &
  \colhead{Comment}   \\
  \colhead{} & 
  \colhead{} &
  \colhead{}  
}
\startdata
\emph{Other Designations}  & \\
KIC 6444896   & \\
KOI 3138   & \\
LSPM J1930+4149   & \\
\Gaia\ DR2 2125699062780742016   & \\
\\
\emph{Basic Information} & \\
Right Ascension & 19:30:00.9006122986  & A \\
Declination & +41:49:49.513849537  & A \\
Proper Motion in RA [\ensuremath{\rm mas\,yr^{-1}}]& -135.842  $\pm$  0.112&A  \\
Proper Motion in Dec [\ensuremath{\rm mas\,yr^{-1}}]& -99.232  $\pm$  0.139&A  \\
Distance to Star~[pc]& 92.5 $\pm$ 0.5 & A\\
\Gaia\ G-magnitude & 16.2682 $\pm$   0.001 & A\\ 
K-magnitude & 12.589 $\pm$   0.026 &  B\\ 
\\
\emph{Stellar Parameters} & \\
Mass, $M_\star$~[$M_\odot$] & \mst\  $\pm$ $ \umst$ & A,B,C \\
Radius, $R_\star$~[$R_\odot$] & \rst\  $\pm$ $ \urst$ & A,B,C \\
Surface Gravity, $\log g_\star$~[cgs] & 5.004  $\pm$ 0.021 & A,B,C \\
Metallicity \metallicity & -0.15  $\pm$ 0.11 & E \\
Effective Temperature, $T_{\rm eff}$ [K] & 3240  $\pm$ $ 61$ & E\\
Luminosity [$L_\odot$] & \lum\  $\pm$  \ulum & C\\
 & & \\
\emph{\thisfirstplanet} & \\
Orbital Period, $P$~[days] & \perplb\  $\pm$ $ \uperplb $ & D \\
Radius Ratio, $R_P/R_\star$ & \rprstb\  $\pm$ $ \urprstb$ & D \\
Scaled semimajor axis, $a/R_\star$  & \arstb\ $\pm~ \uarstb$ & D \\
Orbital inclination, $i$~[deg] & \inclb $ \uinclb$ & D \\
Transit impact parameter, $b$ & \impb $ \uimpb$ & D \\
Transit Duration, $t_{14}$~[hours] & \tdurb \utdurb & D \\
Time of Transit, $t_{t}$~[BJD] & \ttransitb\  $\pm$ \uttransitb & D\\ 
Planet Radius, $R_P$~[\rearth] & \rplb\    $\pm$ $ \urplb$  & A,B,C,D \\
Incident Flux, $S$~[S$_\oplus$] & \fluxb\ $\pm$ \ufluxb  & A,B,C,D,E \\
Equilibrium Temperature, $T_{eq}$~[K] & \teqb\   $\pm$  \uteqb  & A,B,C,D,E,F \\
 & & \\
\emph{\thissecondplanet} &  \\
Orbital Period, $P$~[days] & \perplc\  $\pm$ $\uperplc $ & D \\
Radius Ratio, $R_P/R_\star$ & \rprstc $ \urprstc$ & D \\
Scaled semimajor axis, $a/R_\star$  & \arstc\ $\pm~ \uarstc$ & D \\
Orbital Inclination, $i$~[deg] & \inclc\ $\pm~ \uinclc$ & D \\
Transit Impact parameter, $b$ & \impc\  $\pm$ $ \uimpc$ & D \\
Transit Duration, $t_{14}$~[hours] & \tdurc  \utdurc & D \\
Time of Transit, $t_{t}$~[BJD] & \ttransitc\  $\pm$ \uttransitc & D\\
Planet Radius, $R_P$~[\rearth] & \rplc\ $ \urplc$  & A,B,C,D \\
Incident Flux, $S$~[S$_\oplus$] & \fluxc\ $\pm$ \ufluxc\  & A,B,C,D,E \\
Equilibrium Temperature, $T_{eq}$~[K] & \teqc\   $\pm$   \uteqc  & A,B,C,D,E,F \\
 & & \\
\emph{Other Fit Parameters} &  \\
\bedit{Linear limb darkening parameter [$u_1$]} & \bedit{\uone\  $\pm$  \uuone} & \bedit{G}\\
\bedit{Quadratic limb darkening parameter [$u_2$]} &\bedit{ \utwo\  $\pm$  \uutwo} & \bedit{G}\\
\bedit{Transformed limb darkening parameter 1 [$q_1$]} &\bedit{ \qone\  $\pm$  \uqone} & \bedit{G}\\
\bedit{Transformed limb darkening parameter 2 [$q_2$]} & \bedit{\qtwo\  $\pm$  \uqtwo} & \bedit{G}\\
\bedit{Constant flux offset parameter [$\delta F$]} & \bedit{\offset\  $\pm$  \uoffset} & \bedit{D}\\
& & \\
\enddata
\tablecomments{A: Parameters come from \Gaia\ DR2. B: Parameters come from 2MASS \citep{twomass}. C: Parameters come from empirical relations \citep{mann2015, mann2019}. D: Parameters come from our transit analysis described in Section \ref{lcmodel}.  E: Parameters come from \citet{angelo}. F: Equilibrium temperatures $T_{eq}$ calculated assuming circular orbits, albedo $\alpha$ uniformly distributed between 0 and 0.7, and perfect heat redistribution. $T_{eq}= T_{\rm eff}(1 - \alpha)^{1/4}\sqrt{\frac{R_\star}{2a}}$. \bedit{G: Constrained by an informative prior on $u_1$ and $u_2$ based on model limb darkening parameters \citep{claretbloemen}.}}
\end{deluxetable*}

\section{Statistical Validation}
\label{validation}

Like most stars observed by \Kepler, \thisstar\ is too faint for radial velocity observations to confirm its planets' existence. Instead, we statistically show that \thissecondplanet\ is almost certainly a genuine exoplanet. To do this, we use the \texttt{vespa} software \citep{morton2015}, which implements the methods described by \citet{morton2012} to calculate any given planet candidates's false positive probability (FPP). \texttt{Vespa} uses knowledge of a candidate's orbital characteristics, transit light curve, host stellar parameters, location in the sky, and observational constraints to calculate the relative likelihood that the candidate signal is due to a transiting planet compared to several different false positive scenarios. 

We ran \texttt{vespa} on both \thissecondplanet\ and the previously-validated \thisfirstplanet. We input our well-determined stellar parameters, improved transit light curve, and the high-resolution imaging of \thisstar\ obtained by \citet{angelo}. Like \citet{angelo}, we found that \thisfirstplanet\ is almost certainly a planet, with a false positive probability of $2\times10^{-5}$. The results for \thissecondplanet, on the other hand, were more ambiguous. Due to its slightly shorter transit duration and lower signal-to-noise, it is harder to distinguish the transit light curve of \thissecondplanet\ from that of an extremely grazing eclipsing binary. \texttt{Vespa} reflects this uncertainty by calculating a $\approx$2\% chance that \thissecondplanet\ is an eclipsing binary. However, this calculation does not take into account the presence of a nearby validated planet. If \thisfirstplanet\ is indeed a planet orbiting \thisstar, then \thissecondplanet\ cannot be an eclipsing binary also orbiting \thisstar\ or the system would become dynamically unstable.\footnote{Using the analytic stability criterion from \citet{gladman1993}, we calculate that the total mass of \thisstar\ b and c must be less than about 2 Jupiter masses.} Since \thissecondplanet\ can only be an eclipsing binary if \thisfirstplanet\ does not orbit \thisstar, we multiply the prior for the eclipsing binary scenario in \vespa\ by $1.6\times10^{-5}$ (the probability that \thisfirstplanet\ does not orbit \thisstar). Taking this into account, we find that \thissecondplanet's false positive probability is about $2\times10^{-3}$, well below the thresholds typically applied to consider a planet statistically validated \citep[e.g.,][]{rowe,morton16}. We get a similar result if we apply a ``multiplicity boost'' to \thissecondplanet's false positive probability, to reflect the fact that planets tend to be found in multi-transiting systems more often than false positives, which are more randomly distributed over the stars observed by \Kepler. Applying a multiplicity boost of 10--20 for the DR25 catalog \citep{burke} yields a similar false positive probability of about $1\times10^{-3}$ to $2\times10^{-3}$.




\texttt{Vespa} evaluates almost all astrophysical false positive scenarios, but does not consider false positives due to instrumental artifacts. Most \Kepler\ instrumental false positives are for low signal-to-noise, long-period candidates with only a handful ($\lesssim$~5) \bedit{of} transits \citep{thompson}, though \citet{burke} recently showed that there may be some instrumental false alarm contaminants at short periods (25-100 days) as well. The mechanism behind these short-period false alarms is unknown, but appears only to be significant at very low signal-to-noise ratios (SNR); the examples identified by \citet{burke} all have SNR $\leq$ 8.1. 
\thissecondplanet\ transited 39 times during the \Kepler\ observations, so it does not fall in the main population of long-period \Kepler\ instrumental signals, and is significantly higher SNR (9.3 as calculated by the \Kepler\ pipeline, and 11 in our improved light curve) than the \citet{burke} short-period false positives. We conclude that \thissecondplanet\ is not in a regime where instrumental false positive signals are a serious concern and consider it validated as a \textit{bona fide} exoplanet. 


\section{Discussion}
\label{discussion}

\subsection{\thissecondplanet\ and the Habitable Zone}

\begin{figure*}[htb] 
   \centering
   \includegraphics[width=\linewidth]{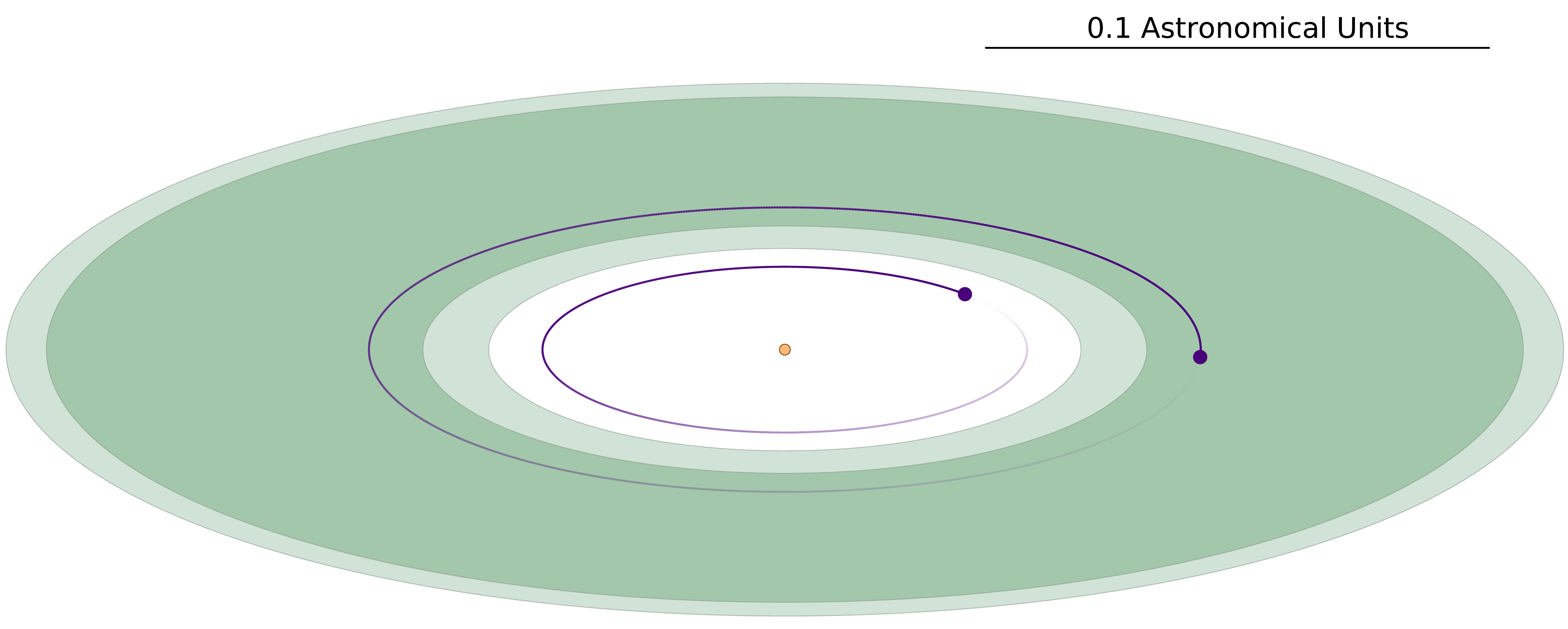} 
   \caption{Diagram of the \thisstar\ system, from the vantage of an observer inclined by 20 degrees from the plane of the system. The host star, \thisstar, is shown to scale in the center of the image, colored as a 3200~K blackbody would appear to the naked eye (see \url{http://www.vendian.org/mncharity/dir3/blackbody/UnstableURLs/bbr_color.html}). The optimistic and conservative habitable zone defined by \citet{kopparapu} are colored in light and dark green, respectively. The orbits of \thisstar\ b and c are shown as faded purple curves. The purple dots at the ends of the orbit curves denoting \thisstar\ b and c are not to scale; the planets' true sizes would be about 4 times smaller than the orbit curve widths. } 
   \label{system}
\end{figure*}

\thissecondplanet\ is an Earth-sized planet orbiting within its host star's habitable zone \citep[as calculated by][under conservative assumptions]{kopparapu}. Figure \ref{system} shows a schematic of the \thisstar\ system along with the location of the circumstellar habitable zone. Though Earth-sized habitable-zone planets are believed to be intrinsically common, they remain difficult to detect, and we only know of a handful today. Based on the NASA Exoplanet Archive Confirmed Planets table\footnote{\url{https://exoplanetarchive.ipac.caltech.edu/cgi-bin/TblView/nph-tblView?app=ExoTbls&config=planets}}, and using \Gaia-based radii from \citet{Berger2018} for Kepler planets, only four transiting planets (TRAPPIST-1 e, f, and g, \citealt{gillon}, and TOI 700 d, \citealt{gilbert20, rodriguez20, suissa20}) and three non-transiting (Proxima Centauri b, \citealt{angladaescude}, Teegarden's Star c, \citealt{teegarden}, and GJ 1061 d \citealt{gj1061}) have radii smaller than 1.25~\rearth, or mass less than 2.0~\mearth, and orbit within their star's conservative habitable zone.\footnote{When published, \Kepler-186 f's radius was estimated to be 1.11$\pm$ 0.14 \rearth\ \citep{quintana}, but \Gaia\ data pushes the planet radius to 1.26 $\pm$ 0.07 \rearth, just above our cutoff.} 


\begin{figure*}[htb!] 
   \centering
   \includegraphics[width=\linewidth]{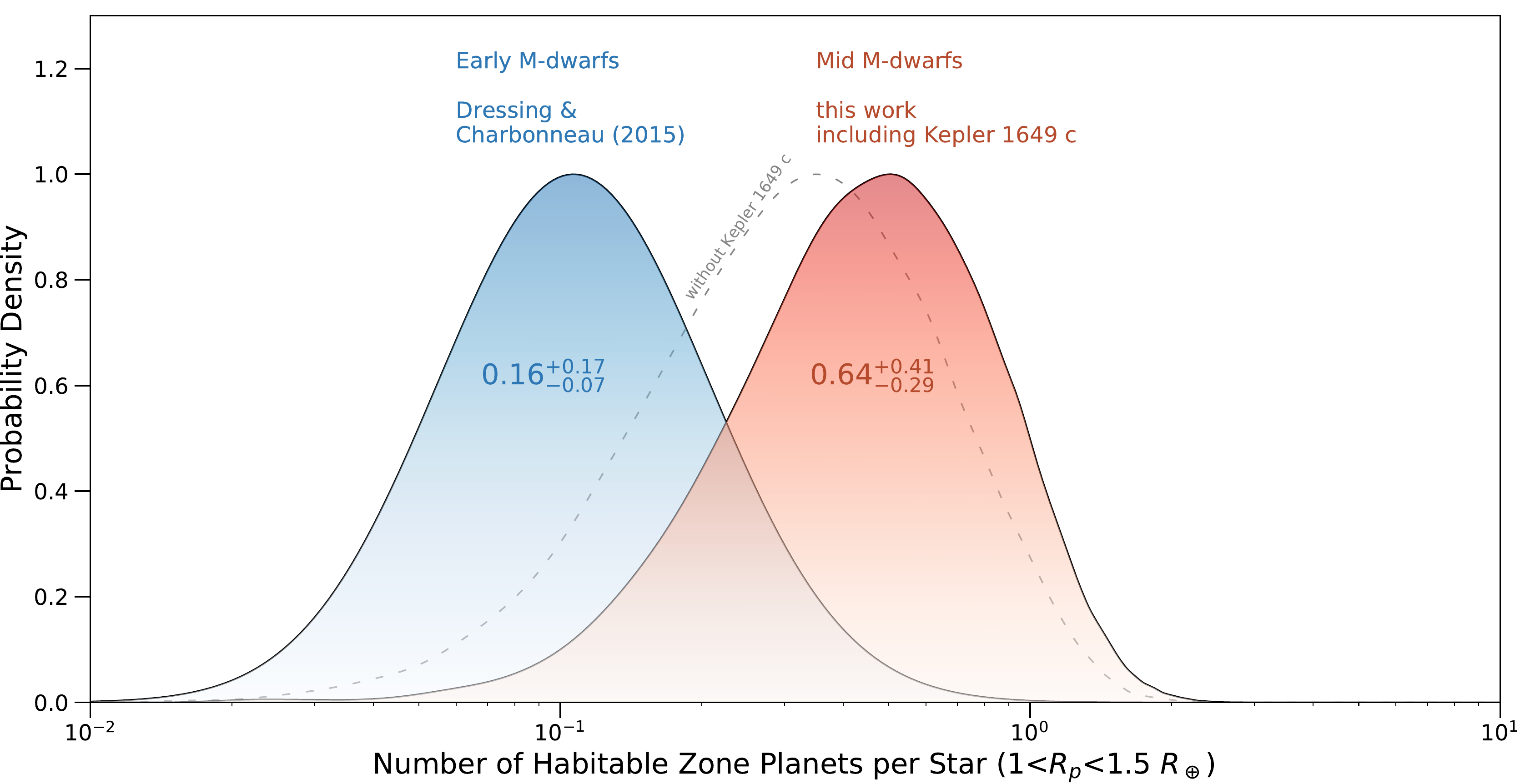} 
   \caption{Posterior probability distributions on the occurrence rate of planets between 1 and 1.5 \rearth\ in size in their stars' conservative habitable zones. We show both the occurrence rate for these planets around early M-dwarf stars (blue distribution) from \citet{dc15}, and mid M-dwarf stars (red distribution), including \thissecondplanet\ in the calculation. We approximated the posterior distribution for early M-dwarf stars by finding (via least squares minimization) a log-normal distribution whose median and 68\% confidence intervals most closely matched (within about 1\%) those reported by \citet{dc15}. We also show the posterior probability distribution for mid M-dwarfs without including \thissecondplanet\ as a grey dashed line. The addition of \thissecondplanet\ slightly increases the measured occurrence rate of small, habitable-zone planets around mid M-dwarfs and adds new evidence suggesting these planets may be more common than those around higher-mass stars. } 
   \label{occurrence}
\end{figure*}

In terms of size and incident \bedit{bolometric} flux, \thissecondplanet\ is a near analog of Earth. Its radius (\rplc\urplc\rearth) is consistent with that of Earth, and the planet receives about 75\% of Earth's incident stellar flux. It seems likely, but not guaranteed, that \thissecondplanet\ has a rocky composition --- hot Earth-sized planets do tend to be rocky \citep{rogers}, but it may be unwise to extrapolate these results to cooler planets like \thissecondplanet. Indeed, though some of \thissecondplanet's bulk parameters are similar to Earth, the planet may not be at all ``Earth-like.'' Many of \thissecondplanet's properties remain uncertain, and planets orbiting M-dwarfs experience a very different environment (an extended era of UV irradiation, tidal locking, etc) from the planets in our own solar system \citep{shields}. 

\subsection{\thisstar\ System Architecture}

Like many other \Kepler\ systems \citep{fabrycky2014}, and especially mid M-dwarfs \citep{muirhead2015}, \thisstar\ hosts multiple close-in transiting planets. Generally, multi-transiting systems are nearly coplanar, with only a small spread in mutual inclinations (\citealt{fabrycky2014}, though not always, see \citealt{kepler108}).  The \thisstar\ system appears to fit this trend, as the inclinations of the two planets are consistent (at the 1$\sigma$ level) with being perfectly coplanar.

The period ratio between \thisstar\ b and c is 2.248250 $\pm$ 0.000013, only 0.08\% inside of a $\sfrac{9}{4}$ period ratio. Often, near-integer period ratios between neighboring planets indicate that the planets were or are in an orbital resonance, but the 9:4 resonance is weak; usually planet pairs are found near stronger resonances like the 2:1 or 3:2 ratios. We therefore suspect that there may be a third planet orbiting between \thisstar\ b and c, forming a chain of 3:2 resonances. If we assume that \thisstar\ b and c are in a three-body Laplace resonance with this hypothetical third planet, its period should be close to 13.029593 days (with an uncertainty of a few minutes). We checked the \Kepler\ light curve for a planet with this orbital period, but found no transits deeper than about 600 ppm. So if there is a third planet orbiting between \thisstar\ b and c, it is either too small to detect (roughly Mars-sized or less, though transit timing variations could hide a somewhat larger planet), or it is misaligned enough to prevent it from transiting. 

\subsection{The Frequency of Habitable Zone Earth-sized Planets around Mid M-dwarfs}

\thissecondplanet\ is the first habitable-zone Earth-sized planet to be found among the mid M-dwarf stars observed by \Kepler. Only about 450 such stars were observed during \Kepler's primary mission \citep{hardegreeullman}, so it is somewhat surprising that \Kepler\ detected a transiting Earth analog in this small sample. We quantified the likelihood that \Kepler\ would find at least one Earth analog around a mid M-dwarf with a Monte Carlo simulation. We generated many ($\approx 10^4$) synthetic planet populations around 461 mid M-dwarfs (with spectral types M3-M5.5) observed by \Kepler\ identified by \citet{hardegreeullman}, assuming planet occurrence statistics from \citet{dc15}. The planets had radii randomly drawn from a uniform distribution between 0.75 and 1.25 \rearth, orbital periods drawn from a log-uniform distribution inside the habitable zone, and inclinations drawn from a uniform distribution in $\cos{(i)}$. We calculated the expected transit signal-to-noise for each favorably-inclined planet, and used the \Kepler\ detection efficiency curve from \citet{Christiansen2017} to determine which simulated planets were detectable by \Kepler. We found only a 3.7\% chance that \Kepler\ would detect an Earth-sized planet in the conservative habitable zone around a mid M-dwarf. 

Given the long, but not insurmountable odds against its detection, it is possible that \thissecondplanet's discovery was just lucky chance, but it is also possible that the intrinsic occurrence rate of such objects we used is incorrect. Our previous calculation assumed an occurrence rate calculated by \citet{dc15} for planets between 1.0 and 1.5 \rearth\ orbiting a sample of mostly early M-dwarf stars. It is plausible that mid M-dwarf stars might form more Earth analogs than their more massive counterparts, thus boosting the chances we would detect such a planet in the \Kepler\ sample. To test this, we calculated the occurrence rates of planets around mid M-dwarf stars observed by \Kepler, while including the newly-rescued \thissecondplanet\ in our calculation. We modeled the population of planets orbiting mid M-dwarf stars (in particular, the sample defined by \citealt{hardegreeullman}) with a joint (un-broken) power law distribution in planet radius and orbital period \citep{burke15, bryson20} and determined power law parameters by exploring a Poisson likelihood function with MCMC. Other than the inclusion of \thissecondplanet, the details of our calculation are identical to that of \bedit{\cite{brysonrnaas}}. 

We note that by including \thissecondplanet\ in our calculations, we make an implicit assumption that the pipeline error leading to \thissecondplanet's incorrect false positive classification was rare and therefore not well represented in the \Kepler\ DR25 vetting completeness experiments. This assumption seems reasonable --- the root cause of \thissecondplanet's incorrect false positive disposition, the large quarter-to-quarter variations in the light curve's photometric scatter, is unusual in \Kepler\ data, only affecting faint stars with high proper motions unknown to the \Kepler\ pipeline. The quarterly variations in \thisstar's light curve are among the worst (95$^{\rm th}$ percentile) of even the faint, high-proper-motion stars in the \Kepler\ mid M-dwarf sample. A visual inspection of the TCEs and the DR25 injection/recovery results for other stars with high quarter-to-quarter variations in photometric scatter showed that no other planet candidates (real or simulated) were rejected on similar grounds to \thissecondplanet. We therefore include \thissecondplanet\ in our calculations with the caveat that if our assumption is incorrect, our occurrence rate may be biased slightly high. 

We integrated the resulting power-law planet occurrence rates over the habitable zones for each star in our sample and calculated an average number of planets per star. Our result is shown in Figure \ref{occurrence}, compared to the \citet{dc15} occurrence rate for such planets around early M-dwarfs. We measure an occurrence rate higher than that of early M-dwarfs, though the statistical significance of this difference is not high. However, when taken together with previous suggestions of increased planet occurrence around mid M-dwarfs \citep{muirhead, hardegreeullman} and the discovery of three Earth-sized planets in the conservative habitable zone of TRAPPIST-1 \citep{gillon}, the detection of \thissecondplanet\ provides further evidence that Earth analogs may be more common around mid M-dwarfs than higher-mass stars. 


\subsection{Human Inspection of Automatically Vetted Signals}


Moving forward, automatic vetting of planet candidates can only become more important as data volume increases and classification techniques improve. There are likely hundreds of undiscovered planets left in K2 data \citep{Dotson2019}, and NASA's TESS satellite produces more light curves and TCEs every month than the entire 10-year \Kepler\ mission (N. Guerrero et al. \textit{submitted}). Expecting humans to keep up with such vast quantities of data is unsustainable, and automatic vetting techniques have already taken the bulk of the triage/vetting workload (\citealt{yu}, N. Guerrero et al. \textit{submitted}).  However, as our rescue of \thissecondplanet\ reinforces, careful human inspection will remain valuable going forward. Even if inspecting each false positive TCE is unfeasible, examining a small but strategically-chosen sample\footnote{\bedit{Some possible subsets of false positives to examine manually could include those around the most likely stars to host detectable transiting planets (like nearby dwarf stars or known transiting planet hosts), those around stars with unusual properties that might trip up automatic classifiers (young stars, very low-mass M-dwarfs, or white dwarfs), and TCEs which just barely missed the cutoff to become planet candidates.}} of targets could help improve automatic methods and enable new discoveries. 


\acknowledgments

We thank Juliette Becker and Courtney Dressing for helpful conversations. AV's work was performed under contract with the California Institute of Technology / Jet Propulsion Laboratory funded by NASA through the Sagan Fellowship Program executed by the NASA Exoplanet Science Institute.

This research has made use of NASA's Astrophysics Data System, the NASA Exoplanet Archive, which is operated by the California Institute of Technology, under contract with the National Aeronautics and Space Administration under the Exoplanet Exploration Program, and the SIMBAD database,
operated at CDS, Strasbourg, France. 

This paper includes data collected by the \Kepler\ mission. Funding for the \Kepler\ mission is provided by the NASA Science Mission directorate. Some of the data presented in this paper were obtained from the Mikulski Archive for Space Telescopes (MAST). STScI is operated by the Association of Universities for Research in Astronomy, Inc., under NASA contract NAS5--26555. Support for MAST for non-HST data is provided by the NASA Office of Space Science via grant NNX13AC07G and by other grants and contracts. This work has made use of data from the European Space Agency (ESA) mission {\it Gaia} (\url{https://www.cosmos.esa.int/gaia}), processed by the {\it Gaia} Data Processing and Analysis Consortium (DPAC, \url{https://www.cosmos.esa.int/web/gaia/dpac/consortium}). Funding for the DPAC
has been provided by national institutions, in particular the institutions participating in the {\it Gaia} Multilateral Agreement.\\

%

\facilities{\Kepler}


\software{IDL Astronomy Library \citep{idlastronomylibrary},  
          \texttt{vespa} \citep{morton12, morton2015}, 
          matplotlib \citep{plt},
          numpy \citep{np}
          }


\vspace{5mm}



\end{document}